\documentclass{article}
\usepackage{arxiv}

\usepackage[utf8]{inputenc}

\title{Inherent Limitations of AI Fairness}
\author{Maarten Buyl\\
Ghent University\\
\texttt{maarten.buyl@ugent.be}
\And
Tijl De Bie\\
Ghent University\\
\texttt{tijl.debie@ugent.be}}
\date{}

\usepackage{xparse, xspace}

\usepackage{graphicx}
\usepackage{svg}

\begin{document}

\maketitle
\begin{abstract}
As the real-world impact of Artificial Intelligence (AI) systems has been steadily growing, so too have these systems come under increasing scrutiny. In response, the study of AI fairness has rapidly developed into a rich field of research with links to computer science, social science, law, and philosophy. Many technical solutions for measuring and achieving AI fairness have been proposed, yet their approach has been criticized in recent years for being misleading, unrealistic and harmful.

In our paper, we survey these criticisms of AI fairness and identify key limitations that are inherent to the prototypical paradigm of AI fairness. By carefully outlining the extent to which technical solutions can realistically help in achieving AI fairness, we aim to provide the background necessary to form a nuanced opinion on developments in fair AI. This delineation also provides research opportunities for non-AI solutions peripheral to AI systems in supporting fair decision processes.
\end{abstract}

\section{Introduction}
In the past, the range of tasks that a computer could carry out was limited by what could be hard-coded by a programmer. Now, recent advances in machine learning make it possible to learn patterns from data such that we can efficiently automate tasks where the decision process is too complex to manually specify. After sensational successes in computer vision and natural language processing, the impact of artificial intelligence (AI) systems powered by machine learning (ML) is rapidly widening towards other domains.

Today, AI is already being used to make high-stakes decisions in areas like predictive policing \cite{larsonHowWeAnalyzed2016}, employment \cite{freireERecruitmentRecommenderSystems2021} and credit scoring \cite{thomasCreditScoringIts2017}. These are highly sensitive applications, in which decisions must be \textit{fair}, i.e. non-discriminatory with respect to an individual's \textit{protected traits} such as gender, ethnicity or religion. Indeed, the principle of non-discrimination has a long history and intersects with many viewpoints: as a value in moral philosophy \cite{biettiEthicsWashingEthics2020}, as a human right \cite{donahoeArtificialIntelligenceHuman2019}, and as a legal protection \cite{wachterWhyFairnessCannot2021}. Evidently, algorithms that play a role in such decision processes should meet similar expectations, which is why \textit{AI fairness} is prominently included as a value in the AI ethics guidelines of many public- and private-sector organizations \cite{jobinGlobalLandscapeAI2019}. Yet AI systems that blindly apply ML are rarely fair in practice, to begin with because training data devoid of undesirable biases is hard to come by \cite{holsteinImprovingFairnessMachine2019}. At the same time, fairness is difficult to hard-code, because it demands nuance and complexity. 

Hence, fairness is both important and difficult to achieve, and as such it is high on the AI research agenda \cite{mehrabininarehSurveyBiasFairness2021}. The dominant approach to AI fairness in computer science is to formalize it as a mathematical constraint, before imposing it upon the decisions of an AI system while losing as little predictive accuracy as possible. For example, assume that we want to design an AI system to score people who apply to enrol at a university. To train this AI system, we may want to learn from data of past decisions made by human administrators in the admission process. Yet if those decisions consistently undervalued women in the past, then any system trained on such data is likely to inherit this bias. In the prototypical fair classification setting, we could formalize fairness by requiring that men and women are, on average, admitted at equal proportions. A simple technical AI fairness approach to meet this constraint is to uniformly increase the scores given to women. The resulting decisions may then be considered fair, and little accuracy is likely to be lost. 

Yet this approach rests on simplistic assumptions. For instance, it assumes that fairness can be mathematically formalized, that we collect gender labels on every applicant, that everyone fits in a binary gender group, and that we only care about a single axis of discrimination. The validity of such assumptions has recently been challenged, raising concerns over technical interventions in AI systems as a panacea for ensuring fairness in their application 
\cite{selbstFairnessAbstractionSociotechnical2019,cooperEmergentUnfairnessAlgorithmic2021, luSubvertingMachinesFluctuating2022}.

Despite this criticism, the idea of tackling fairness through mathematical formalism remains popular. To an extent, this is understandable. Large-scale data collection on humans making high-impact decisions could enable us to study biases in the allocation of resources with a level of statistical power that was previously infeasible. If we would be able to properly mitigate biases in AI systems, then we may even want fair AI to replace these human decision makers, such that the overall fairness of the decision process is improved. At the very least, technical solutions in fair AI seem to promise pragmatic benefits that are worth considering.

Therefore, the aim of this article is in estimating how far (technical) AI fairness approaches can go in truly measuring and achieving fairness by outlining what \textit{inherently} limits it from doing so in realistic applications. With this lens, we survey criticisms of AI fairness and distill eight such inherent limitations. These limitations result from shortcomings in the assumptions on which AI fairness approaches are built. Hence, they are considered fundamental, practical obstacles and we will not frame them as research challenges that can be solved within the strict scope of AI research. Rather, our aim is to provide the reader with a disclaimer for the ability of fair AI approaches to address fairness concerns. By carefully delineating the role that it can play, technical solutions for AI fairness can continue to bring value, though in a nuanced context. At the same time, this delineation provides research opportunities for non-AI solutions peripheral to AI systems, and mechanisms to help fair AI fulfill its promises in realistic settings.

\textit{Remark} In what follows, we start from the technical perspective on AI fairness such that we can test its inherent limits and their relation with socio-technical, legal and industry challenges. Related work takes a different approach by arguing that, since fairness is fundamentally a socio-technical concept, we should anyhow not expect technical tools to be sufficient in pursuing fairness \cite{selbstFairnessAbstractionSociotechnical2019,wongSeeingToolkitHow2023}. The significance of technical tools should instead be found in the informational role they play in service of a wider socio-technical discussion \cite{abebeRolesComputingSocial2020}. The present paper is written from a different perspective, thus corroborating this conclusion.

\section{Limitations of AI Fairness}
\begin{figure}
	\centering
	\includegraphics[width=0.8\linewidth]{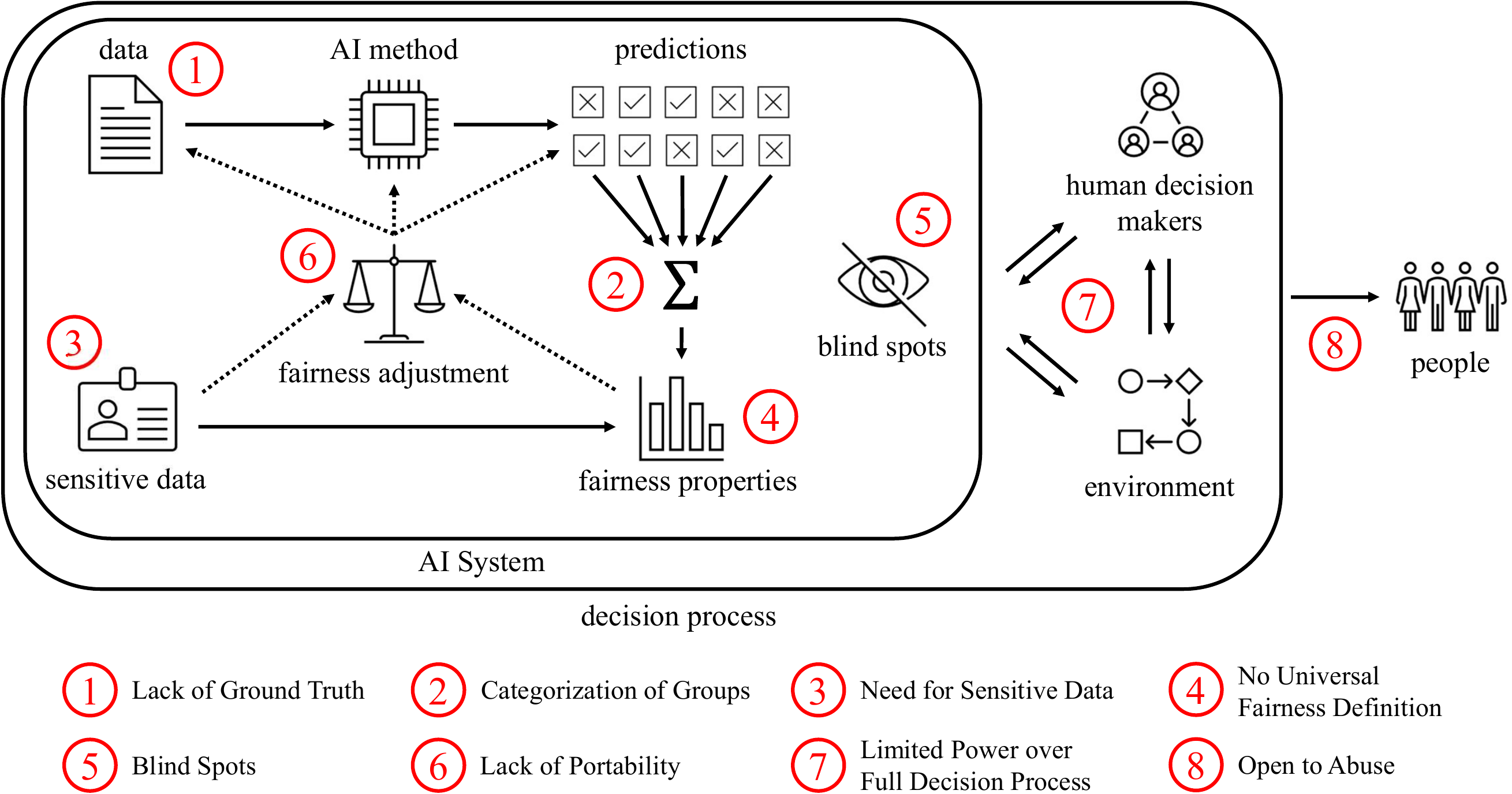}
	\caption{A prototypical fair AI system. Each limitation affects a different component of the full decision process.}
	\label{fig:overview}
	\vskip-0.2in
\end{figure}

In Fig.~\ref{fig:overview}, we provide an overview of the prototypical AI fairness solution \cite{mehrabininarehSurveyBiasFairness2021}. In this setting, an AI method learns from data, which may be biased, to make predictions about individuals. Task-specific fairness properties are computed by categorizing individuals into \textit{protected groups} like \textit{men} and \textit{women} and then comparing aggregated statistics over the predictions for each group. Without adjustment, these predictions are assumed to be biased, because the algorithm may inherit bias from the data, and because the algorithm is probably imperfect and may then make worse mistakes for some of the groups. Fairness adjustment is done using \textit{preprocessing} methods that attempt to remove bias from the data, \textit{inprocessing} methods where modifications are made to the learning algorithm such that discriminatory patterns are avoided, and \textit{postprocessing} methods that tweak the predictions of a potentially biased learning algorithm.

We consider eight inherent limitations of this prototypical fair AI system, which each affect its different components and levels of abstraction as illustrated in Fig.~\ref{fig:overview}. To start, we observe that bias in the data results in biased approximations of the \textit{ground truth}, leading to unfair conclusions about the performance and fairness properties of the AI system (Sec.~\ref{sec:truth}). Fairness measurements are also problematic because they involve distinguishing people into groups (Sec.~\ref{sec:cat}) and require \textit{sensitive data} of individuals (Sec.~\ref{sec:sens_data}) to do so. In fact, there is generally no universal definition of fairness in exact terms (Sec.~\ref{sec:general}). Avoiding \textit{any} discrimination is anyhow unrealistic, as it demands that the possible sources of bias are well-understood without any blind spots (Sec.~\ref{sec:blind}). Even if fairness can be adequately elaborated for one setting, the same solutions will often not be portable to another (Sec.~\ref{sec:port}). More generally, AI systems are only a component of a larger decision process, e.g. where biases can arise in the interaction with human decision makers and the environment, that is no longer within the scope of the AI (Sec.~\ref{sec:power}). In this larger scope, we conclude that AI fairness can be abused, through negligence or malice, to worsen injustice (Sec.~\ref{sec:abuse}).

\subsection{Lack of Ground Truth}\label{sec:truth}
%Machine learning researchers are used to distrusting datasets. %Though it is generally agreed that dataset samples are largely informative of reality, 
%To begin, noise in the features and any labels is more common than not. %At times, a limited amount of noise is even welcomed, because it provides an incentive to not only learn from the training dataset, but also generalize to unseen data. 
%Assumptions about noise in the data can therefore be found throughout machine learning literature. This model of noise in the dataset usually assumes that, given enough data, the noise is averaged out. In classification settings, this means that we can even use noisy output labels as an \textit{unbiased} estimate of the \textit{ground truth}, so long as we evaluate enough data points. %This paragraph can probably be removed.
%\tijl{Ik denk ook dat deze paragraaf misschien beter verdwijnt. Het is een open deur dat data ruis bevat, en de rest is niet zo belangrijk lijkt me. Ik laat de beslissing aan jou. (Begin volgende paragraaf moet "Yet, " dan ook verdwijnen.)}

%Yet, 
As is now widely understood by the AI fairness community, datasets that involve humans are often plagued by a range of biases that may lead to discrimination in algorithmic decisions \cite{mehrabininarehSurveyBiasFairness2021}. Importantly, this bias may prevent an unbiased estimation of the ground truth, i.e. the prediction that should have been made for an individual. It is then difficult to make unbiased statements about the performance of an algorithm \cite{cooperEmergentUnfairnessAlgorithmic2021}. For example in predictive policing, when trying to predict whether someone will commit a crime, the dataset may use `arrested' as the output label that must be predicted. Yet when one group is policed more frequently, their arrest rates may not be comparable to an advantaged group \cite{sureshFrameworkUnderstandingSources2021}, meaning it cannot be assumed to be an unbiased estimate of the ground truth `committed a crime' \cite{dresselAccuracyFairnessLimits2018}. This raises concerns whether predictive policing can ever be ethical after long histories of biased practices \cite{richardsonDirtyDataBad2019}.

%Thus, also the accuracy of the algorithm is not comparable between these different groups.

In general, the lack of a ground truth is a significant problem for fair AI, as it often depends on the availability of the ground truth to make statements about fairness. For instance, a large disparity between false positive rates formed the basis of criticism against the COMPAS algorithm, which was found to incorrectly predict high recidivism risk for black defendants more often than for white defendants \cite{larsonHowWeAnalyzed2016}. However, lacking a ground truth to compute those rates in an unbiased way, one cannot even measure algorithmic fairness in this setting \cite{friedlerImPossibilityFairness}. Caution is thus warranted in the interpretation of any metrics where ground truth should be used to compute them. This holds for overall performance metrics such as accuracy, but also for many fairness statistics. 

%Ending this discussion, we point out the contrast with the AIA proposal, which asserts that the dataset should be "complete, high quality, ...". -> note: there is a lot of debate around this phrasing already online.

%Ultimately, the bias of a dataset can only be evaluated in a subjective manner.

%data is reflection of reality. if reality is biased, then data bias is always viewed through a sociopolitical lens

%Should probably not conduct more surveillance on under-represented groups.

\subsection{Categorization of Groups}\label{sec:cat}
Bias has been considered in algorithms since at least 1996 \cite{friedmanBiasComputerSystems1996}. Yet, it was the observations that biases are found throughout real datasets and then reproduced by data-driven AI systems that led to the rapid development of AI fairness research \cite{barocasFairnessMachineLearning2019}. AI systems are typically evaluated and optimized with a formal, mathematical objective in mind, and so this field demands formalizations of fairness that somehow quantify discriminatory biases in exact terms.  

Assuming the AI system is meant to classify or score individuals, a strong formalization of fairness is the notion of \textit{individual} fairness, which asserts that "any two individuals who are similar with respect to a particular task should be classified similarly" \cite{dworkFairnessAwareness2012}. Though principled, such definitions require an unbiased test to assess whether two individuals are indeed similar. However, if such a test were readily available, then we could directly use that test to construct an unbiased classifier. Developing a test for strong definitions of individual fairness is thus equally hard as solving the initial problem of learning a fair classifier.

The vast majority of AI literature is instead concerned with the more easily attainable notion of \textit{group fairness}, which requires that any two protected groups should, on average, receive similar labels \cite{berkConvexFrameworkFair2017, mehrabininarehSurveyBiasFairness2021}. Group fairness expresses the principles of individual fairness \cite{binnsApparentConflictIndividual2020} by looking at the sum of discrimination towards an entire group, rather than individual contributions. Though this increased statistical power makes group fairness much more practical to measure and satisfy, it comes with its own problems which we discuss next.

\begin{figure}
	\centering
	\includegraphics[width=0.9\linewidth]{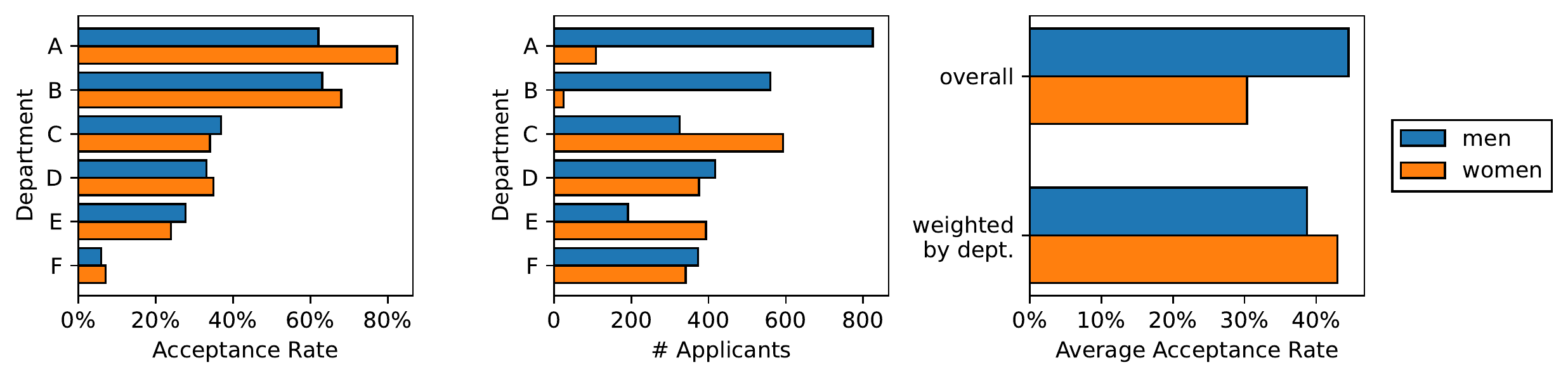}
	\caption{UC Berkeley admissions statistics for men and women. \textbf{Left:} acceptance rates. \textbf{Middle:} number of applicants. \textbf{Right:} average acceptance rate, either overall or weighted by the total number of applicants (of both groups) for each department.}
	\label{fig:berkeley}
	\vskip-0.2in
\end{figure}

\subsubsection{Simpson's Paradox}\label{sec:simpson}
An obvious problem with group fairness is that, through aggregation, some information will be lost. To illustrate, we discuss the example of the UC Berkeley admission statistics from the fall of 1973 \cite{freedmanStatisticsFourthInternational2007}. These statistics showed that the overall acceptance rate for men was higher (44\%) than for women (35\%), suggesting there was a pattern of discrimination against the latter. However, when we consider the acceptance rates separately per department\footnote{Note that most departments were omitted in the example.} in Fig.~\ref{fig:berkeley}, there does not seem to be such a pattern. In fact, except for departments C and D, the acceptance rate is reportedly higher for women. From this perspective, it could be argued that women are mostly favoured. The fact that conclusions may vary depending on the granularity of the groups is referred to as \textit{Simpson's Paradox}.

The paradox arises from the difference in popularity between the departments. Women more commonly applied to departments with low acceptance rates (e.g. departments C through F), which hurt their \textit{overall} average acceptance rate. Yet when taking the \textit{weighted} average by accounting for the total number of applicants of both men and women, we see that women score higher. As noted by Wachter et al. \cite{wachterWhyFairnessCannot2021}, the point of this example is not that some forms of aggregation are better than others. Rather, we should be wary that aggregation, such as the kind that is found in group fairness, may influence the conclusions we draw.

\subsubsection{Fairness Gerrymandering}
An especially worrisome property of group fairness is that, by looking at averages over groups, it allows for some members of a disadvantaged group to receive poor scores (e.g. due to algorithmic discrimination), so long as other members of the same group receive high enough scores to compensate. Such situations are considered \textit{fairness gerrymandering} \cite{kearnsPreventingFairnessGerrymandering2018} if specific subgroups can be distinguished who are systematically disadvantaged within their group.

\begin{figure}
	\centering
	\includegraphics[width=0.9\linewidth]{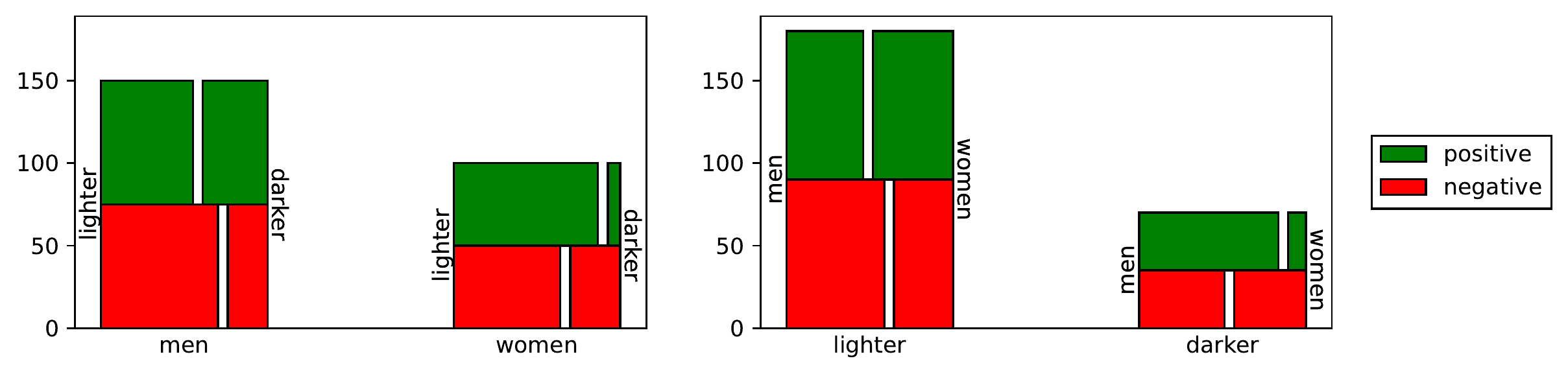}
	\caption{Toy example where fairness along two axes separately does not imply fairness along both axes simultaneously. Each vertical bar shows the number of positives and negatives for a group seen along the axis of either gender (\textbf{left}) or skin tone (\textbf{right}) separately. Each bar is also split in two such that the proportions of for each intersectional subgroup (e.g. \textit{darker-skinned women}) is visible.}
	\label{fig:intersect}
	\vskip-0.2in
\end{figure}

Group fairness measures may thus be hiding some forms of algorithmic discrimination \cite{kasyFairnessEqualityPower2021}. The toy example in Fig.~\ref{fig:intersect} illustrates this. It is constructed such that \textit{men} and \textit{women} receive positive decisions at equal average rates (here 50\%), just like when we view the groups of \textit{lighter-skinned} and \textit{darker-skinned} people separately. A system giving such scores might therefore be seen as fair, because group fairness measures no discrepancy based on gender or skin tone. However, this per-axis measurement hides the discrimination that may be experienced at the intersection of both axes. In the case of Fig.~\ref{fig:intersect}, lighter-skinned men and darker-skinned women proportionally receive fewer positive predictions than the other intersectional groups. Because darker-skinned women are also in the minority, they receive even fewer positive predictions proportionally (through similar effects as in Sec.~\ref{sec:simpson}). Consider then that in the real world, we often do see significant discrimination along separate axes. The discrimination experienced by those at the intersection may then be far worse. Indeed, empirical evidence shows for example that darker-skinned women often face the worst error rates in classification tasks \cite{buolamwiniGenderShadesIntersectional2018, ghoshCharacterizingIntersectionalGroup2021}. 

%\begin{table}[]
%\begin{tabular}{l|cc|cc} 
%& \multicolumn{2}{c|}{lighter-skinned} & %\multicolumn{2}{c}{darker-skinned} \\
%& pop. size  & avg. score & pop. size & avg. score \\\hline
%men & 100 & 48 \% & 50  & 54\% \\
%women   & 80   & 52.5\%  & 20  & 40\% \\
%\end{tabular}
%\caption{Toy example of a situation where there are equal average scores for men and women, and for light-and darker-skinned people. Though the scores may be fair per axis of protected groups, the intersectional subgroup of darker-skinned women clearly receives far worse scores than any other subgroup.}
%\label{tab:intersect}
%\end{table}

The field of \textit{intersectional fairness} is concerned with addressing the (magnified) discrimination experienced by those at the intersection of protected groups \cite{buolamwiniGenderShadesIntersectional2018}. A straightforward approach is to directly require fairness for all combinations of subgroups. While this addresses some of the concerns for the example in Fig.~\ref{fig:intersect}, it is not a realistic solution. Such definitions of intersectional fairness give rise to a number of subgroups that grows exponentially with the number of axes of discrimination \cite{kearnsPreventingFairnessGerrymandering2018, ghoshCharacterizingIntersectionalGroup2021}, thereby losing the statistical power we were hoping to gain with group fairness.

\subsubsection{Discretization Concerns}
Socio-technical views of intersectional fairness also criticise the assumption that people can neatly be assigned to groups at all. This assumption helps in mapping legal definitions of discrimination to a technical context, yet it thereby also inherits the issues that result from rigid identity categories in law \cite{hoffmannWhereFairnessFails2019}. In fact, most protected characteristics resist discretization \cite{luSubvertingMachinesFluctuating2022}. Gender and sexual orientation are unmeasurable and private \cite{rakovabogdanaWhereResponsibleAI2021}, disability is highly heterogeneous \cite{buylTacklingAlgorithmicDisability2022}, and ethnicity labels are subjective \cite{kasirzadehUseMisuseCounterfactuals2021} and may be reported differently depending on the time and place \cite{luSubvertingMachinesFluctuating2022}. 

While the fear of gerrymandering in group fairness encourages highly granular (intersections of) identity groups, an `overfitting' of identity categorization should thus also be avoided. It must be better understood how machine learning, which aims at generalization, can model such individual nuances.

%note bias introduced by algorithms? 

\subsection{Need for Sensitive Data}\label{sec:sens_data}
Fair AI research is fuelled by the use of sensitive data, i.e. data on people's traits that are protected from discrimination. Almost all methods require such data to measure and mitigate bias \cite{mehrabininarehSurveyBiasFairness2021}, as explicitly shown by Žliobaitė and Custers \cite{zliobaiteUsingSensitivePersonal2016}. Indeed, exact fairness definitions as discussed in Sec.~\ref{sec:cat} involve the categorization of people into groups, which requires us to know, for every person, which group they actually identify with (if any).

Yet it is hardly a viable assumption that sensitive data is generally available for bias mitigation purposes in the real world. Clearly, the collection of such highly personal data conflicts with the ethical principle of privacy \cite{vealeFairerMachineLearning2017}, despite the fact that privacy and fairness are often considered joint objectives in ethical AI guidelines \cite{jobinGlobalLandscapeAI2019}. Fairness should not blindly be given priority here, since disadvantaged groups may prefer to avoid disclosing their sensitive data due to distrust caused by structural discrimination in the past \cite{andrusDemographicReliantAlgorithmicFairness2022}. For example, outing the sexual orientation or gender identity of queer individuals can lead to emotional distress and serious physical or social harms \cite{tomasevFairnessUnobservedCharacteristics2021}.

Indeed, practitioners report that individuals are often hesitant to share sensitive data, most frequently because they do not trust that it will be in their benefit \cite{andrusWhatWeCan2021}. Privacy and availability concerns are thus consistently cited as significant obstacles to implementing fair AI in practice \cite{lawDesigningToolsSemiAutomated2020, holsteinImprovingFairnessMachine2019, rakovabogdanaWhereResponsibleAI2021}. Sensitive data has received special protection in the past, such as in European data protection law \cite{edwardsSlaveAlgorithmWhy2017}. In stark contrast, the European Commission's recently proposed AI Act \cite{europeancommissionEURLex52021PC0206EURLex} now includes a provision that specifically allows for the processing of sensitive data for the purpose of ensuring "the bias monitoring, detection and correction in relation to high-risk AI systems".

\subsection{No Universal Fairness Definition}\label{sec:general}
So far, we have discussed mathematical definitions of fairness in simple terms, i.e. as constraints that compare simple statistics of different demographic groups. These constraints are motivated through the idea that the AI system is distributing resources, which ought to be done in a \textit{fair} way. However, does this mean that each group should receive an \textit{equal} amount of resources, or should we try to be \textit{equitable} by taking the particular need, status and contribution of these groups into account \cite{youngEquityTheoryPractice1995}? This irresolution has led to a notoriously large variety of fairness notions. For just the binary classification setting, one survey from 2018 \cite{vermaFairnessDefinitionsExplained2018} lists 20 different definitions, each with their own nuance and motivations. For example, \textit{statistical parity} \cite{dworkFairnessAwareness2012} requires \textit{equality} in the rate at which positive predictions are made. On the other hand, \textit{equalised odds} \cite{hardtEqualityOpportunitySupervised2016} requires the \textit{false positive} and \textit{false negative rates} to be equal. It therefore allows for different rates at which groups receive positive decisions, so long as the amount of mistakes is proportionally the same. Both notions are defensible depending on the context, but they are also generally incompatible: if the base rates differ across groups (i.e. one group has a positive label in the data more often than another), then statistical parity and equalised odds can only be met simultaneously in trivial cases \cite{kleinbergInherentTradeOffsFair2016}.

In fact, many notions have shown to be mathematically incompatible in realistic scenarios \cite{friedlerImPossibilityFairness,pleissFairnessCalibration2017}. This has led to controversy, e.g. in the case of the COMPAS algorithm. As mentioned in Sec.~\ref{sec:truth}, it displayed a far higher false positive rate in predicting recidivism risk for black defendants than white defendants \cite{larsonHowWeAnalyzed2016}. Yet the COMPAS algorithm was later defended by noting that it was equally calibrated for both groups, i.e. black and white defendants that were given the same recidivism risk estimate indeed turned out to have similar recidivism rates \cite{floresFalsePositivesFalse}. Later studies proved that such a balance in calibration is generally incompatible with a balance in false positive and negative rates, except in trivial cases \cite{kleinbergInherentTradeOffsFair2016,pleissFairnessCalibration2017}.
From a technical point of view, these results imply that there cannot be a single definition of fairness that will work in all cases. Consequently, technical tools for AI fairness should be flexible in how fairness is formalized, which greatly adds to their required complexity.

Moreover, it may not even be possible to find a `most relevant' fairness definition for a specific task. Indeed, human-subject studies show that humans generally do not reach a consensus on their views of algorithmic fairness \cite{grgic-hlacaDimensionsDiversityHuman2022, harrisonEmpiricalStudyPerceived2020, starkeFairnessPerceptionsAlgorithmic2021, albachRoleAccuracyAlgorithmic2021}. For example, it appears that people's perceptions of fairness are heavily influenced by whether they personally receive a favourable outcome \cite{wangFactorsInfluencingPerceived2020}. The `most relevant' fairness definition may therefore be the product of a constantly shifting \cite{binnsItReducingHuman2018} compromise resulting from a discussion with stakeholders. It is also noted that these discussions may be hampered by a lack of understanding of technical definitions of fairness \cite{sahaMeasuringNonExpertComprehension2020}.

\subsection{Blind Spots}\label{sec:blind}
%In our discussion on previous limitations, we have shown
It will now be clear that fairness is inherently difficult to formalize and implement in AI methods. We thus stressed the need for nuance and iteration in addressing biases. However, such a process assumes that possible biases are already anticipated or discovered. Businesses have raised concerns that, though they have processes and technical solutions to tackle bias, they can only apply them to biases that have already been foreseen or anticipated \cite{lawDesigningToolsSemiAutomated2020}. Without an automated process to find them, they are limited to following hunches of where bias might pop up. Holstein et al. \cite{holsteinImprovingFairnessMachine2019} cite one tech worker saying "you’ll know if there’s fairness issues if someone raises hell online". Though well-designed checklists may help to improve the anticipation of biases \cite{madaioCoDesigningChecklistsUnderstand2020}, it is safe to assume that some blind spots will always remain. %Calling an AI system `fair' should thus always come with the disclaimer that `unknown unknowns' have not been solved. 
Since unknown biases are inherently impossible to measure, we cannot always make definitive guarantees about fairness in their presence. The fairness of AI systems should thus constantly be open to analysis and criticism, such that new biases can quickly be discovered and addressed.

In fact, some biases may only arise \textit{after} deployment. For example, the data on which an algorithm is trained may have different properties than the data on which it is evaluated, because the distribution of the latter may be continuously changing over time \cite{kohWILDSBenchmarkIntheWild2021}. Similarly, any fairness measurements taken during training may not be valid after deployment \cite{damourFairnessNotStatic2020}. The AI Act proposed by the European Commission \cite{europeancommissionEURLex52021PC0206EURLex} would therefore rightly require that high-risk AI systems are subjected to post-market monitoring, in part to observe newly arising biases. The fairness properties of an AI system should thus continuously be kept up-to-date \cite{hardtAmazonSageMakerClarify2021}. 

\subsection{Lack of Portability}\label{sec:port}
As we have argued, fairness should be pursued for a particular AI system after carefully elaborating the assumptions made (e.g. those in Sec.~\ref{sec:truth} and Sec.~\ref{sec:cat}) and after finding compromise between different stakeholders and viewpoints (e.g. because fairness is difficult to formalize as argued in Sec.~\ref{sec:general}). However, this means that our desiderata for a fair AI system become highly situation-specific. As pointed out by Selbst et al. \cite{selbstFairnessAbstractionSociotechnical2019}, this limits the portability of fairness solutions across settings, even though portability is a property that is usually prized in computer science. Consequently, the flexibility of fair AI methods that we wanted in Sec.~\ref{sec:general} may not be achievable by simply choosing from a zoo of portable fairness solutions.

In industry, there is already empirical evidence that `off-the-shelf' fairness methods have serious limits. For example, fairness toolkits such as \textit{AIF360}, \textit{Aequitas} and \textit{Fairlearn} offer a suite of bias measurement, visualization and mitigation algorithms. However, though such toolkits can be a great resource to learn about AI fairness \cite{richardsonFairnessPracticePractitionerOriented2021}, practitioners found it hard to actually adapt them to their own model pipeline or use case \cite{leeLandscapeGapsOpen2021}. 
%linear fairness assumptions? 

%\subsection{Fairness is Expensive}
%investment costs: need to involve stakeholders, adapt inflexible models, do research, post-deployment monitoring, ...

%short-term accuracy loss

\subsection{Limited Power over Full Decision Process}\label{sec:power}
Fair AI papers often start by pointing out that algorithms are increasingly replacing human decision-makers. Yet decisions in high-risk settings such as credit scoring, predictive policing, or recruitment are expected to meet ethical requirements. Such decision processes should thus only be replaced by algorithms that meet or exceed similar ethical requirements, such as fairness. However, it is unrealistic to assume that decision processes will be fully automated in precisely those high-risk settings that motivate fair AI in the first place. This is because fully automated AI systems are not trusted to be sufficiently accurate and fair \cite{binnsItReducingHuman2018, knowlesSanctionAuthorityPromoting2021}. The EU's General Data Protection Regulation even specifically grants the right to \textit{not} be subject to fully automated decision-making, in certain circumstances \cite{allenArtificialIntelligenceRight2020}. 
%\cite{xuHumanJudgesEra2022}: "Therefore, it is important to make it clear that judicial artificial intelligence is only a helper of human judges, not a stand-in."

In most high-risk settings, algorithms only play a supportive role and the final decision is subject to human oversight \cite{greenFlawsPoliciesRequiring2022}. However, unfairness may still arise from the interaction between the algorithm and the larger decision process (e.g. including human decision-makers) that is outside its scope \cite{selbstFairnessAbstractionSociotechnical2019}. For example, a study conducted by Green and Chen \cite{greenPrinciplesLimitsAlgorithmintheLoop2019} asked participants to give a final ruling on defendants in a pretrial setting. After being presented with an algorithmic risk assessment, participants tended to (on average) assign a higher risk to black defendants than the algorithm. The reverse was true for white defendants.

Hence, in an unjust world, it is meaningless to talk about the fairness of an algorithm's decisions without considering the wider socio-technological context, in which an algorithm is applied \cite{fazelpourAlgorithmicFairnessNonideal2020}. Instead, it is more informative to measure the overall fairness properties of a decision process, of which an algorithm may only be a small component among human decision makers and other environmental factors \cite{cruzcortesLocalityTechnicalObjects2022}.

\subsection{Open to Abuse}\label{sec:abuse}
Like most technology, solutions for algorithmic fairness are usually evaluated with the understanding that they will be used in good faith. Yet the sheer complexity of fairness may be used as a cover to avoid fully committing to it in practice. Indeed, opportunistic companies implementing ethics into their AI systems may resort to \textit{ethics-shopping}, i.e. they may prefer and even promote interpretations of fairness that align well with existing (business) goals \cite{floridiTranslatingPrinciplesPractices2019}. Though they may follow organizational `best practices' by establishing ethics boards and collecting feedback from a wide range of stakeholders, an actual commitment to moral principles may mean doing more than what philosophers are allowed to do within corporate settings \cite{biettiEthicsWashingEthics2020}. Fundamentally, solutions towards ethical AI have a limited effect if a deviation from ethics has no consequences \cite{hagendorffEthicsAIEthics2020}.

%Like most technology, solutions for algorithmic fairness are usually motivated and validated with the understanding that they will be used in good faith. Yet the sheer complexity of fairness may be used as a pretext to avoid spending too much effort on addressing it. Indeed, it is a common provision in law that regulations should only be met insofar as that is technologically feasible. If the bar is not set high enough, then solutions that aim to achieve fairness through excessive simplification could be abused to claim that system is made `fair enough' \cite{biettiEthicsWashingEthics2020}. This danger is not limited to technical solutions, but could also occur when organizations are satisfied with affirming fairness as part of the principles they follow, without seriously putting them to practice \cite{hagendorffEthicsAIEthics2020}. 
%terms of service-achtige dingen in apps, cookies. 
%iets meer referenties, maar speculatie is doelbewust 
%teaching to the test -> verschuilen achter metrieken die het beste uitkomt. eerder opportunisme. 

The complexity of fairness may not only lead to obscure commitments towards it; its many degrees of freedom may also be abused to create or intensify injustice. In Sec.~\ref{sec:truth}, we argued that ground truth labels are often unavailable, lending power to whomever chooses the proxy that is used instead. Moreover, as discussed in Sec.~\ref{sec:cat}, the groups in which people are categorized grants the power to conceal discrimination against marginalized subgroups. Following our discussion in Sec.~\ref{sec:sens_data} on the need for sensitive data to measure fairness, we here also warn that this necessity could motivate further surveillance, in particular on those already disadvantaged \cite{cooperEmergentUnfairnessAlgorithmic2021}. Overall, significant influence is also afforded in deciding how fairness is actually measured, as a universal definition may not exist (see Sec.~\ref{sec:general}). 

\section{Conclusion}
It is evident from our survey of criticism towards fair AI systems that such methods can only make guarantees about fairness based on strong assumptions that are unrealistic in practice. Hence, AI fairness suffers from inherent limitations that prevent the field from accomplishing its goal on its own. Some technical limitations are inherited from modern machine learning paradigms that expect reliable estimates of the ground truth and clearly defined objectives to optimize. Other limitations result from the necessity to measure fairness in exact quantities, which requires access to sensitive data and a lossy aggregation of discrimination effects. The complexity of fairness means that some forms of bias will always be missed, and that every elaboration of fairness is highly task-specific. Moreover, even a perfectly designed AI system often has limited power to provide fairness guarantees for the full decision process, as some forms of bias will remain outside its scope. Finally, the extensive automation of high-stakes decision processes with allegedly fair AI systems entails important risks, as the complexity of fairness opens the door to abuse by whomever designs them.

These inherent limitations motivate why AI fairness should not be considered a panacea. Yet we stress that also the many benefits of AI fairness must not be overlooked, since it can remain a valuable tool as part of broader solutions. In fact, many of the limitations we identify and assumptions we question are not only inherent to fair AI, but to the ethical value of fairness in general. The study of AI fairness thus forces and enables us to think more rigorously about what fairness really means to us, lending us a better grip on this elusive concept. In short, fair AI may have the potential to make society more fair than ever, but it needs critical thought and outside help to make it happen.

\section*{Acknowledgements}
This research was funded by the ERC under the EU's 7th Framework and H2020 Programmes (ERC Grant Agreement no. 615517 and 963924), the Flemish Government (AI Research Program), the BOF of Ghent University (PhD scholarship BOF20/DOC/144), and the FWO (project no. G0F9816N, 3G042220).

\bibliographystyle{plain}
\bibliography{references}
\end{document}